# Short-Term Precision and Least Significant Change of 3D-DXA Cortical and Trabecular Proximal Femur Measurements Across Hologic DXA Scanner Models

**Authors:** Asier Muñoz[1]*, Thomas L. Nickolas[2], Laurent Maïmoun[3,4], Polyzois Makras[5], Jorge Malouf[6], Xavier Nogués[7,8], Ludovic Humbert[1]

**Affiliation:**

[1]3D-Shaper Medical, Barcelona, Spain

[2]Division of Bone and Mineral Diseases, Department of Medicine, Washington University School of Medicine, St. Louis, MO, United States

[3]Service de Médecine Nucléaire, Hôpital Lapeyronie, CHU Montpellier, Montpellier, France

[4]Physiologie et Médecine Expérimentale du Cœur et des Muscles (PhyMedEx), INSERM, CNRS, Université de Montpellier, Montpellier, France

[5]Department of Endocrinology and Diabetes and Department of Medical Research, 251 Hellenic Air Force & VA General Hospital, Athens, Greece

[6]Mineral Metabolism Unit, Grup Creu Groga, Mataró, Spain

[7]Hospital del Mar Research Institute, Centro de Investigación Biomédica en Red de Fragilidad y Envejecimiento Saludable (CIBERFES), Barcelona, Spain

[8]Internal Medicine Department, Hospital del Mar, Universitat Pompeu Fabra, Barcelona, Spain

* Corresponding author:
Asier Muñoz
3D-Shaper Medical, Rambla Catalunya 53-55, 4th floor, 08007 Barcelona, Spain
Email: asier.munoz@3d-shaper.com

## Abstract

3D-DXA provides volumetric and compartment-specific hip measurements from standard DXA scans. However, for reliable longitudinal interpretation, establishing scanner-specific short-term precision data is essential. This study assessed the short-term precision of 3D-DXA-derived parameters using repeated acquisitions from five clinical Hologic DXA scanner datasets. Duplicate hip DXA acquisitions were collected at five clinical centers using two Horizon Wi scanners, two Horizon A scanners, and one Discovery W scanner. Each subject was scanned twice with complete repositioning between acquisitions. Conventional total hip and femoral neck aBMD were obtained using APEX software, and 3D-Shaper software was used to derive integral vBMD, trabecular vBMD, and cortical sBMD. Precision error was expressed as RMS-SD and RMS-CV, and LSC was calculated at the 95% confidence level. For total hip aBMD, absolute LSC values ranged from 0.018 to 0.037 g/cm$^2$. Femoral neck aBMD showed higher error, with LSC values ranging from 0.031 to 0.049 g/cm$^2$. For 3D-DXA parameters, absolute LSC values ranged from 9.955 to 17.859 mg/cm$^3$ for integral vBMD, 8.662 to 18.487 mg/cm$^3$ for trabecular vBMD, and 4.444 to 8.875 mg/cm$^2$ for cortical sBMD. Precision errors for 3D-Shaper parameters varied moderately across the five clinical datasets. Short-term precision of 3D-Shaper-derived measurements was broadly comparable with previously published precision data. The LSC values reported here can support interpretation of longitudinal changes in integral vBMD, trabecular vBMD, and cortical sBMD on Hologic scanners.

## 1. Introduction

Clinical guidelines widely recommend dual-energy X-ray absorptiometry (DXA) for osteoporosis diagnosis and monitoring because it is accessible, relatively inexpensive, noninvasive, and associated with very low radiation exposure. Conventional DXA measures areal bone mineral density (aBMD) from a two-dimensional (2D) projection of the mineralized bone signal along the X-ray path and cannot separately quantify the cortical and trabecular compartments, which can lose bone at different rates [1] and respond differently to pharmacological treatment [2,3]. Quantitative computed tomography (QCT) provides three-dimensional (3D) compartmental assessment, but its higher cost and radiation exposure make it impractical for routine clinical use. To overcome the limitations of standard 2D imaging without the clinical barriers of adopting new imaging modalities, 3D-DXA technology has been developed, currently available for clinical use through the 3D-Shaper® software (3D-Shaper Medical, Barcelona, Spain). This technology generates a subject-specific 3D model of the proximal femur by registering a statistical shape and density model onto a standard hip DXA scan [4,5], enabling separate analysis of cortical and trabecular bone from routine clinical images.

Any DXA-derived measurement used for monitoring should be interpreted in the context of its short-term precision, which defines the least significant change (LSC), the minimum change required to distinguish true biological or treatment-related variation from the measurement variability introduced by repeated scan acquisition, patient repositioning, image analysis, and scanner-related factors. Conventional DXA precision has been widely studied across GE Healthcare (Madison, WI, USA) and Hologic systems (Marlborough, MA, USA) [6–11]. In contrast, the precision literature for 3D-DXA is still developing. Early work evaluated one Hologic Discovery W and one GE Lunar iDXA system [12], and a subsequent study has expanded data for GE Healthcare Prodigy and iDXA devices [13]. Short-term precision data for 3D-DXA are also available for Discovery A devices [14], but evidence across multiple Hologic scanner models remains limited. Furthermore, to the best of our knowledge, the short-term precision of parameters derived from the recently regulatory-cleared clinical version of 3D-Shaper [15] has not been reported for Hologic systems.

The aim of this study was to evaluate the short-term precision of 3D-Shaper in five datasets spanning three Hologic scanner models. Specifically, the primary objective of this study was to determine the short-term precision and LSC of 3D-Shaper-derived integral vBMD, trabecular vBMD, and cortical sBMD from repeated hip DXA acquisitions across Hologic scanner models. Secondary objectives were to compare precision across scanner models and with conventional DXA measurements, and to evaluate whether subject characteristics, scan mode, mean value, or clinical center influenced scan-rescan variability. These validations are critical given the software's global adoption for clinical use.

## 2. Materials and methods

### 2.1. Subjects and DXA measurements

Short-term precision was evaluated using repeated hip DXA acquisitions collected prospectively at five clinical centers equipped with Hologic scanners: Horizon Wi at Hospital del Mar (Barcelona, Spain), Horizon Wi at 251 Hellenic Air Force & VA General Hospital (Athens, Greece; hereafter 251 HAF Hospital), Horizon A at Hôpital Lapeyronie (Montpellier, France), Horizon A at Washington University School of Medicine (St. Louis, MO, USA; hereafter Washington University), and Discovery W at Hospital de la Santa Creu i Sant Pau (Barcelona, Spain; hereafter Hospital Sant Pau). Eligible subjects were adults referred for routine bone health assessment, with a minimum eligibility age of 20 years. Individuals with surgical implants were excluded. Each subject underwent two hip acquisitions on the corresponding scanner, with complete repositioning between scans. Total hip and femoral neck aBMD were obtained using APEX® software (Hologic, Marlborough, MA, USA). Daily quality control and calibration were performed according to the manufacturer's instructions.

### 2.2.3D-DXA analysis

3D-DXA analyses were performed using 3D-Shaper® software, version 2.14 (3D-Shaper Medical, Barcelona, Spain). In brief, the software generates a subject-specific 3D model of proximal femur geometry and density for the hip DXA scan using statistical shape and density modeling combined with 2D-3D image registration [4]. The software provides three 3D-derived measurements: integral vBMD, trabecular vBMD, and cortical sBMD, at the total hip.

## 2.3.Statistical analysis

Statistical analyses were performed using Microsoft Excel 2021. Subject characteristics and DXA-derived measurements were summarized by clinical center using descriptive statistics calculated from the first acquisition of each subject. Short-term in vivo precision was assessed from duplicate acquisitions according to ISCD recommendations [16]. Precision error was reported in absolute units using the root mean square standard deviation (RMS-SD) and in relative terms using the root mean square coefficient of variation (RMS-CV). The LSC at the 95% confidence level was calculated as 2.77 × precision error and reported in absolute units and as a percentage.

Exploratory analyses were performed to assess whether scan-rescan variability was associated with subject characteristics, scan mode, or clinical center. Absolute scan-rescan error was calculated as the absolute difference between duplicate acquisitions and was used as the outcome for these exploratory analyses. Associations with age, BMI (body mass index), and mean value (the average of the two duplicate acquisitions) were first assessed using Pearson correlation coefficients. Multivariable ordinary least-squares linear regression models were then fitted separately for each DXA-derived parameter, with absolute scan-rescan error as the dependent variable and age, BMI, sex, scan mode (Fast Array versus Array), and mean value as independent variables. Age models were adjusted for mean value, sex, and clinical center, with age scaled per 10 years. BMI models were adjusted for mean value, age, sex, and clinical center, with BMI scaled per 5 $kg/m^2$. Washington University was excluded from BMI correlation and regression analyses because height and weight were not available for the full cohort. Sex models were adjusted for mean value, age, and clinical center. Because every center used a single, fixed scan mode, scan mode was fully confounded with clinical center. Scan mode models were therefore adjusted for mean value, age, sex, and scanner model instead of clinical center. Mean value models were adjusted for age, sex, and clinical center. To further distinguish scanner model effects from center-related variability, additional models compared centers that used the same scanner model: 251 HAF Hospital versus Hospital del Mar for Horizon Wi, and Washington University versus Hôpital Lapeyronie for Horizon A. These models were adjusted for mean value, age, and sex.

Regression coefficients were reported with 95% confidence intervals and *p*-values. All exploratory model *p*-values, including the age, BMI, sex, scan mode, mean value, and within-scanner center comparison models, were corrected for multiple comparisons using the Benjamini-Hochberg procedure, applied separately within each family of five parameter-specific tests. Both unadjusted and Benjamini-Hochberg-adjusted *p*-values are reported.

## 3. Results

### 3.1.Subject characteristics and DXA-derived measurements

Table 1 provides a descriptive summary of the cohorts for each clinical center, including demographic characteristics, anthropometric variables, DXA aBMD values, and 3D-DXA parameters.

**Table 1.** Subject characteristics, DXA scanner models, and DXA and 3D-DXA measurements across clinical centers.

| | **Hospital del Mar** | **251 HAF Hospital** | **Hôpital Lapeyronie** | **Washington University** | **Hospital Sant Pau** |
|---|---|---|---|---|---|
| **Subject characteristics** | | | | | |
| **N** | 31 | 34 | 50 | 58 | 33 |
| **Sex** | 25 women, 6 men | 23 women, 11 men | 38 women, 12 men | 53 women, 5 men | 27 women, 6 men |
| **Age, years** | 64 (13.4) [25-82] | 65 (10.4) [46-83] | 43 (11.6) [21-62] | 50 (14.1) [27-76] | 69 (10.0) [52-91] |
| **Height, cm** | 159.6 (10.2) [146.0-185.0] | 166.8 (9.7) [154.0-189.0] | 164.7 (7.8) [150.0-184.0] | not available | 155.9 (7.9) [143.5-172.0] |
| **Weight, kg** | 66.1 (16.1) [41.0-110.0] | 71.4 (14.8) [43.0-92.0] | 79.8 (14.3) [49.0-109.0] | not available | 64.9 (13.3) [43.8-97.6] |
| **BMI, kg/m²** | 25.8 (5.5) [17.7-45.8] | 25.5 (4.0) [17.4-31.8] | 29.4 (4.7) [19.5-40.4] | not available | 26.7 (5.1) [19.1-39.3] |
| **Femur side** | 1 right, 30 left | 34 left | 50 left | 58 left | 33 left |
| **Scanner information** | | | | | |
| **Hologic scanner** | Horizon Wi | Horizon Wi | Horizon A | Horizon A | Discovery W |
| **Scan mode** | Array | Fast Array | Fast Array | Array | Array |
| **DXA and 3D-DXA measurements[a]** | | | | | |
| **Total hip aBMD, g/cm²** | 0.802 (0.149) [0.562-1.141] | 0.854 (0.131) [0.628-1.227] | 1.004 (0.137) [0.692-1.376] | 0.970 (0.158) [0.655-1.394] | 0.787 (0.129) [0.545-1.134] |
| **Femoral neck aBMD, g/cm²** | 0.642 (0.121) [0.442-0.942] | 0.717 (0.137) [0.424-1.011] | 0.865 (0.143) [0.586-1.199] | 0.804 (0.148) [0.553-1.220] | 0.622 (0.120) [0.360-1.049] |
| **Integral vBMD, mg/cm³** | 285.1 (55.4) [187.8-394.8] | 287.6 (46.5) [202.5-402.1] | 357.3 (59.3) [209.0-468.5] | 340.7 (59.3) [222.1-461.5] | 260.1 (51.9) [179.8-417.9] |
| **Trabecular vBMD, mg/cm³** | 153.4 (37.1) [89.6-229.8] | 159.8 (36.0) [92.7-254.0] | 209.1 (47.1) [101.8-306.5] | 199.0 (46.0) [114.2-315.2] | 138.3 (36.8) [75.6-238.2] |
| **Cortical sBMD, mg/cm²** | 146.0 (30.6) [100.9-214.7] | 149.4 (20.9) [114.1-203.9] | 176.3 (22.7) [122.0-227.0] | 164.3 (23.4) [115.1-208.7] | 135.7 (23.1) [83.5-193.4] |

Quantitative variables are presented as mean (standard deviation) [minimum–maximum]. [a]DXA and 3D-DXA values were calculated from the first scan of each subject. BMI, body mass index; aBMD, areal bone mineral density; vBMD, volumetric bone mineral density; sBMD, surface bone mineral density.

### 3.2.Short-term precision

Short-term precision results are presented in Table 2. For conventional DXA measurements, total hip aBMD generally showed better precision than femoral neck aBMD. RMS-SD values ranged from 0.006 to 0.014 g/cm$^2$ for total hip aBMD and from 0.011 to 0.018 g/cm$^2$ for femoral neck aBMD, corresponding to absolute LSC values of 0.018–0.037 g/cm$^2$ and 0.031–0.049 g/cm$^2$, respectively. The highest total hip aBMD errors were observed at 251 HAF Hospital, while the highest femoral neck aBMD errors were observed at 251 HAF Hospital and Hospital del Mar.

For 3D-DXA parameters, absolute RMS-SD values ranged from 3.594 to 6.447 mg/cm$^3$ for integral vBMD, from 3.127 to 6.674 mg/cm$^3$ for trabecular vBMD, and from 1.604 to 3.204 mg/cm$^2$ for cortical sBMD. Hospital Sant Pau showed the lowest precision errors.

**Table 2.** Short-term scan-rescan precision of DXA and 3D-DXA measurements by clinical center.

| Clinical Center | RMS-SD | LSC, absolute | RMS-CV, % | LSC, relative, % |
|---|---|---|---|---|
| **Hospital del Mar** | | | | |
| **Total hip aBMD, g/cm²** | 0.010 | 0.028 | 1.28 | 3.55 |
| **Femoral neck aBMD, g/cm²** | 0.018 | 0.049 | 2.69 | 7.45 |
| **Integral vBMD, mg/cm³** | 6.447 | 17.859 | 2.23 | 6.18 |
| **Trabecular vBMD, mg/cm³** | 6.674 | 18.487 | 4.86 | 13.45 |
| **Cortical sBMD, mg/cm²** | 3.204 | 8.875 | 2.22 | 6.14 |
| **251 HAF Hospital** | | | | |
| **Total hip aBMD, g/cm²** | 0.014 | 0.037 | 1.56 | 4.32 |
| **Femoral neck aBMD, g/cm²** | 0.016 | 0.043 | 2.27 | 6.28 |
| **Integral vBMD, mg/cm³** | 5.546 | 15.362 | 1.99 | 5.51 |
| **Trabecular vBMD, mg/cm³** | 6.002 | 16.626 | 3.47 | 9.62 |
| **Cortical sBMD, mg/cm²** | 3.085 | 8.545 | 2.11 | 5.85 |
| **Hôpital Lapeyronie** | | | | |
| **Total hip aBMD, g/cm²** | 0.009 | 0.025 | 0.96 | 2.66 |
| **Femoral neck aBMD, g/cm²** | 0.013 | 0.035 | 1.58 | 4.37 |
| **Integral vBMD, mg/cm³** | 5.356 | 14.837 | 1.51 | 4.18 |
| **Trabecular vBMD, mg/cm³** | 4.612 | 12.776 | 2.31 | 6.40 |
| **Cortical sBMD, mg/cm²** | 2.529 | 7.004 | 1.42 | 3.94 |
| **Washington University** | | | | |
| **Total hip aBMD, g/cm²** | 0.006 | 0.018 | 0.68 | 1.89 |
| **Femoral neck aBMD, g/cm²** | 0.014 | 0.038 | 1.69 | 4.69 |
| **Integral vBMD, mg/cm³** | 4.627 | 12.816 | 1.42 | 3.94 |
| **Trabecular vBMD, mg/cm³** | 4.851 | 13.438 | 2.81 | 7.80 |
| **Cortical sBMD, mg/cm²** | 2.595 | 7.189 | 1.57 | 4.35 |
| **Hospital Sant Pau** | | | | |
| **Total hip aBMD, g/cm²** | 0.010 | 0.029 | 1.35 | 3.73 |
| **Femoral neck aBMD, g/cm²** | 0.011 | 0.031 | 2.05 | 5.68 |
| **Integral vBMD, mg/cm³** | 3.594 | 9.955 | 1.46 | 4.05 |
| **Trabecular vBMD, mg/cm³** | 3.127 | 8.662 | 2.25 | 6.22 |
| **Cortical sBMD, mg/cm²** | 1.604 | 4.444 | 1.29 | 3.58 |

RMS-SD and absolute LSC are reported in the units indicated in the parameter name; RMS-CV and relative LSC are reported as percentages. LSC was calculated as 2.77 × precision error. aBMD, areal bone mineral density; vBMD, volumetric bone mineral density; sBMD, surface bone mineral density; RMS-SD, root mean square standard deviation; RMS-CV, root mean square coefficient of variation; LSC, least significant change.

### 3.3. Factors associated with scan-rescan variability

Exploratory analyses were performed to evaluate whether subject characteristics, scan mode, mean value, or center-related factors contributed to differences in short-term precision error. In unadjusted Pearson correlation analyses, BMI, age, and mean value showed no significant associations with absolute scan-rescan error across conventional DXA or 3D-DXA parameters (Supplementary Figure S1). After adjustment, neither age, BMI, sex, nor the mean value was significantly associated with absolute scan-rescan error for any DXA or 3D-DXA parameter (Supplementary Figure S2, Panels A-C, E), before or after correction for multiple comparisons. Scan mode (Fast Array versus Array) was also assessed; Fast Array mode was associated with a higher absolute scan-rescan error for total hip aBMD before correction ($p = 0.018$), but this association did not remain significant after correction for multiple comparisons ($p_{adj} = 0.092$, Supplementary Figure S2, Panel D). No other DXA or 3D-DXA parameter showed a significant association with scan mode, before or after correction.

Additional models compared centers using the same scanner model: 251 HAF Hospital versus Hospital del Mar for Horizon Wi, and Washington University versus Hôpital Lapeyronie for Horizon A (Supplementary Figure S3). For Horizon Wi, no significant adjusted center differences were observed for any conventional DXA or 3D-DXA parameter. For Horizon A, Washington University showed lower adjusted absolute scan-rescan error than Hôpital Lapeyronie for total hip aBMD before correction ($p = 0.026$), but this difference did not remain significant after correction for multiple comparisons ($p_{adj} = 0.128$). No significant differences were observed for femoral neck aBMD, integral vBMD, trabecular vBMD, or cortical sBMD, before or after correction.

## 4. Discussion

This study reports short-term precision data for conventional DXA and 3D-DXA measurements from repeated hip acquisitions across five clinical Hologic scanner datasets. These precision estimates support interpretation of longitudinal changes in 3D-DXA parameters relative to repeated-acquisition error. In addition, the reported scanner-specific LSC values provide clinicians with practical thresholds for distinguishing true biological changes from measurement variability in clinical practice.

For conventional DXA measurements, total hip aBMD generally showed lower precision error than femoral neck aBMD. Absolute LSC values ranged from 0.018 to 0.037 g/cm² for total hip aBMD and from 0.031 to 0.049 g/cm² for femoral neck aBMD (Table 2). These ranges are broadly consistent with previous Hologic precision studies reporting total hip LSC values of approximately 0.021–0.031 g/cm² and femoral neck LSC values of approximately 0.026–0.057 g/cm² [6,7,10–12,14]. They are also comparable to values reported for GE Healthcare Prodigy and iDXA systems [6–9,13,17]. The differences observed between Hologic datasets should not be attributed to scanner model alone, since controlled comparisons between Hologic Horizon and Discovery systems have reported only small and non-significant differences in hip aBMD precision [10]. Untested cohort characteristics and center-specific procedures, including technician training for repositioning and scan analysis, may also contribute to differences in scan-rescan variability between datasets.

Precision was compared mainly using absolute LSC values, since absolute values are recommended over relative (%) values for precision assessment, particularly when baseline measurements differ between subjects or centers [18]. This is consistent with our data: absolute scan-rescan error showed no significant association with mean value for any DXA or 3D-DXA parameter (Supplementary Figure S1, Panels K–O), indicating that error did not scale with the underlying baseline value. Consequently, expressing a constant absolute error as a percentage introduces apparent differences in precision that reflect baseline value differences between subjects or centers rather than true differences in measurement variability. Relative LSC values were also reported for comparison with ISCD-recommended precision criteria. These remained below the ISCD minimum acceptable threshold for total hip aBMD (5%) [19]. For femoral neck aBMD, Hospital del Mar and 251 HAF Hospital showed the highest relative LSC values. However, only the Hospital del Mar Horizon Wi dataset exceeded the ISCD minimum acceptable threshold of 6.9% [19], whereas the 251 HAF Hospital Horizon Wi dataset remained below this threshold (Table 2). The consistently higher femoral neck aBMD errors across all datasets may partly reflect the greater sensitivity of this region to repositioning, femoral rotation, ROI placement, and site-specific acquisition or analysis procedures, consistent with previous studies showing higher precision error for femoral neck aBMD than for total hip aBMD [6–8,10,11,17].

For 3D-DXA parameters, absolute LSC values ranged from 9.955 to 17.859 mg/cm$^3$ for integral vBMD, from 8.662 to 18.487 mg/cm$^3$ for trabecular vBMD, and from 4.444 to 8.875 mg/cm$^2$ for cortical sBMD (Table 2). The lowest errors were observed at Hospital Sant Pau, closely matching previous data reporting LSC values of 10.39 mg/cm$^3$, 9.64 mg/cm$^3$, and 6.25 mg/cm$^2$ for the same parameters [12]. In a previous study using a Discovery A scanner, LSC values estimated from reported precision errors were 11.163 mg/cm$^3$ for integral vBMD, 10.886 mg/cm$^3$ for trabecular vBMD, and 6.039 mg/cm$^2$ for cortical sBMD [14], which are within the range of values reported in the current study. The values reported in our study using Hologic scanners were also broadly comparable to GE Healthcare Prodigy and iDXA data, where reported LSC values ranged from 8.280 to 11.466 mg/cm³ for integral vBMD, from 8.511 to 10.828 mg/cm³ for trabecular vBMD, and from 5.862 to 6.626 mg/cm² for cortical sBMD [13]. Hospital del Mar and 251 HAF Hospital showed the highest absolute precision errors for all three 3D-Shaper parameters, following the pattern observed for femoral neck aBMD. Both centers used a Horizon Wi scanner, but whether these differences are attributable to the scanner model alone cannot be concluded from the available data and study design. Even with the higher errors observed at these two centers, the LSC ranges for integral vBMD, trabecular vBMD, and cortical sBMD remained within the range of previously published 3D-DXA precision data [12–14].

Additional analyses showed that neither age, BMI, sex, nor the mean value was a significant independent predictor of absolute scan-rescan error after adjustment for relevant covariates, before or after correction for multiple comparisons (Supplementary Figure S2 A-C, E). Notably, mean age differed appreciably between centers, ranging from 43 years at Hôpital Lapeyronie and 50 years at Washington University to 64-69 years at the remaining three centers (Table 1); however, our models indicate that this between-center age difference is unlikely to explain the observed variability in scan-rescan precision (Supplementary Figure S2 A). For scan mode, total hip aBMD had higher absolute scan-rescan error with Fast Array mode before correction ($p$ = 0.018), but this association did not remain significant after correction for multiple comparisons ($p_{adj}$ = 0.092, Supplementary Figure S2 D). This is consistent with previous work in a Hologic Discovery A system reporting no significant effect of scan mode on short-term precision of BMD [20]. Center comparisons within the same scanner model showed no significant differences between the two Horizon Wi datasets for any conventional DXA or 3D-DXA parameter (Supplementary Figure S3 A), while for Horizon A, Washington University showed lower absolute scan-rescan error than Hôpital Lapeyronie for total hip aBMD before correction ($p$ = 0.026), but this difference did not remain significant after correction for multiple comparisons ($p_{adj}$ = 0.128, Supplementary Figure S3 B). Overall, no single tested covariate accounted for a dominant share of scan-rescan variability.

Some limitations should be acknowledged. Horizon Wi and Horizon A were each evaluated in two independent clinical centers, whereas Discovery W was represented by a single center-scanner dataset. Scanner model and center-specific effects therefore cannot be fully separated. The within-scanner comparisons provide some insight into center-related variability, but individual contributions from acquisition protocol, patient repositioning, technician experience, and local analysis procedures cannot be isolated, and unmeasured patient characteristics beyond age, BMI, and sex may also have contributed to the observed variability. Analysis software version also varied within some centers, but the resulting subgroups were too small or imbalanced to support a reliable analysis of its effect. In addition, only three Hologic scanner models were included, and the results should not be generalized to all Hologic systems or other DXA manufacturers. However, the primary aim was to establish practical LSC values for 3D-DXA parameters acquired on Hologic scanners under routine clinical conditions, rather than to isolate scanner-specific effects. Furthermore, this study evaluated same-day scan-rescan precision and therefore does not capture additional sources of variability encountered during long-term clinical follow-up. Despite these limitations, the reported LSC values provide a practical framework for interpreting longitudinal changes in 3D-DXA measurements acquired on these systems. Local confirmation of precision before implementing 3D-DXA at a new center or with a different scanner model remains a decision for individual users, who can compare their results against the LSC values reported here.

## 5. Conclusion

This study reports short-term precision values for 3D-DXA hip measurements across five Hologic scanner datasets. The resulting LSC values provide practical thresholds for interpreting whether longitudinal changes in integral vBMD, trabecular vBMD, and cortical sBMD exceed measurement error. Finally, these data support the use of 3D-Shaper-derived measurements for longitudinal follow-up in routine clinical practice and future interventional studies.

## 6. Acknowledgments

The authors thank the study participants for their contribution.

## 7. Funding sources

This research did not receive any specific grant from funding agencies in the public, commercial, or not-for-profit sectors.

## 8. Declaration of interest

A.M. is an employee of 3D-Shaper Medical. T.L.N. serves on the scientific advisory board for Amgen, Micot Pharma, and Pathalys, and has received speaker fees from Intas Pharma. L.H. is an employee and shareholder of 3D-Shaper Medical. The remaining authors declare no competing interests.

## Supplementary material

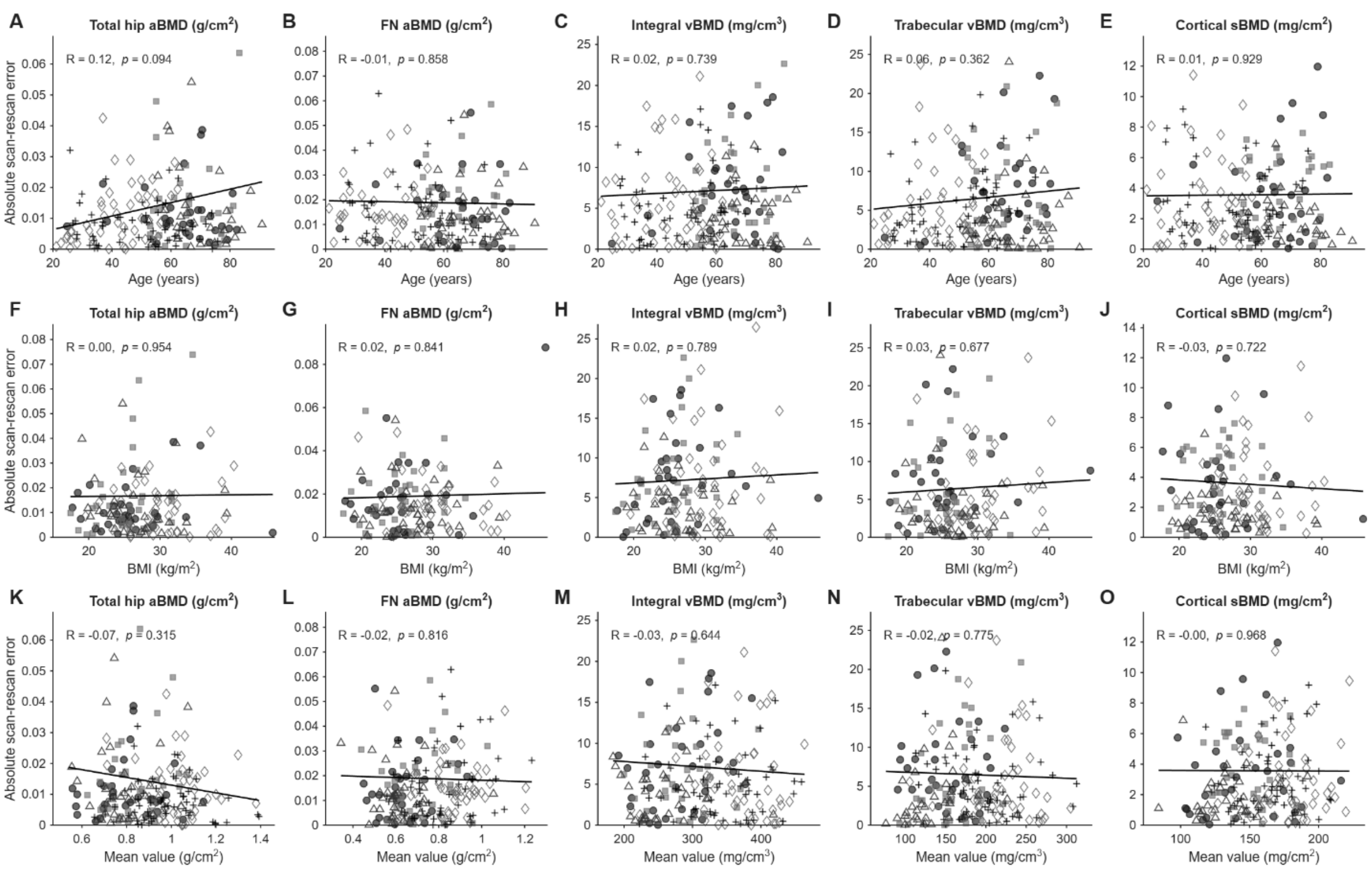

**Supplementary Figure S1.** Associations between subject characteristics and absolute scan-rescan errors. The points represent the individual scan-rescan differences for each parameter; different symbols indicate the clinical centers. Black lines represent unadjusted linear trends, and the displayed R and *p*-values correspond to Pearson correlation analyses. **(A – E)** Age vs. total hip aBMD, femoral neck aBMD, integral vBMD, trabecular vBMD, and cortical sBMD, respectively. **(F – J)** BMI vs. the same parameters, respectively. **(K – O)** Mean value vs. the same parameters, respectively.

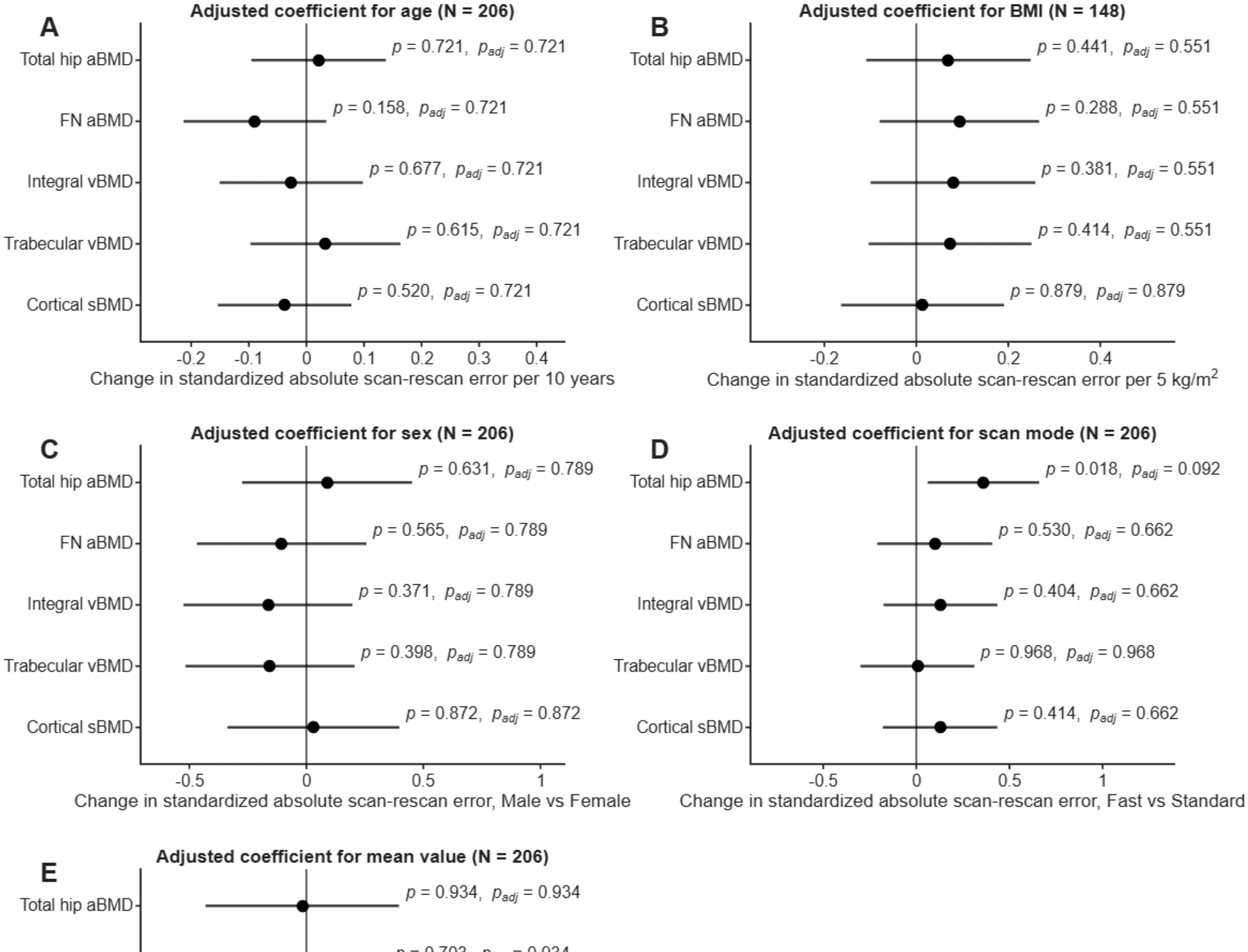

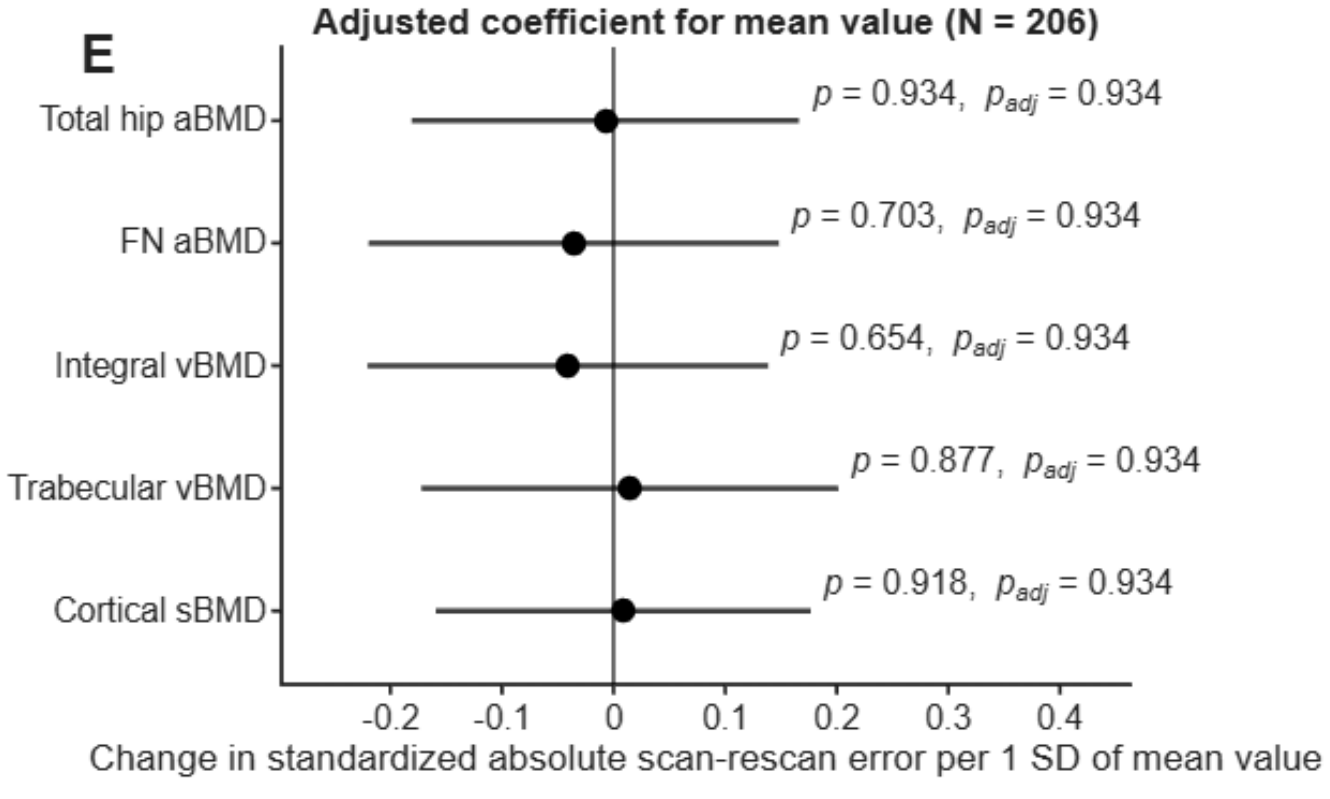

**Supplementary Figure S2.** Adjusted associations between subject characteristics and absolute scan-rescan error. Multivariable regression models were fitted separately for each DXA and 3D-DXA parameter. **(A)** Age, per 10-year increase. **(B)** BMI, per 5 kg/m² increase. **(C)** Sex, Male relative to Female. **(D)** Scan mode, Fast Array relative to Array. **(E)** Mean value, per 1 SD increase in the parameter's own level. For visualization, regression coefficients and 95% confidence intervals were standardized by the standard deviation of absolute scan-rescan error within each parameter, allowing comparison across parameters measured on different scales. Points indicate standardized regression coefficients, and horizontal lines indicate standardized 95% confidence intervals. Raw ($p$) and Benjamini-Hochberg-adjusted ($p_{adj}$) $p$-values are shown for each parameter, corrected within each panel's family of five tests.

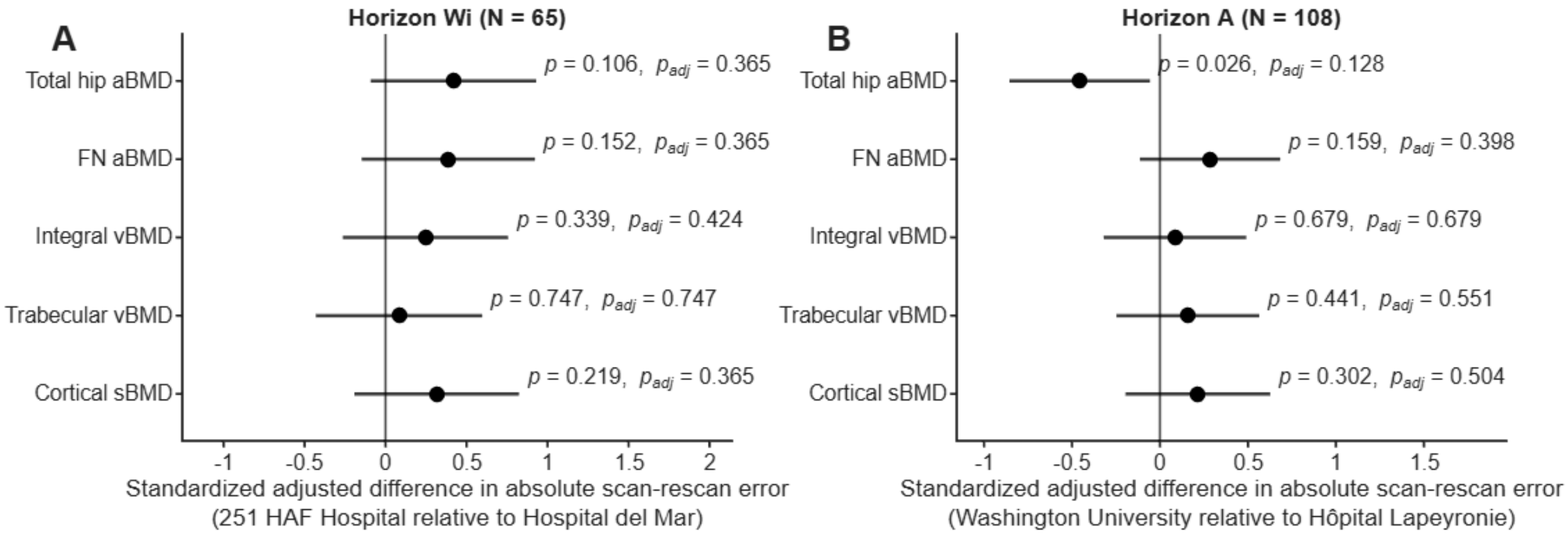

**Supplementary Figure S3.** Adjusted center differences in absolute scan-rescan error among datasets acquired with the same scanner model. Multivariable regression models were fitted separately for each DXA and 3D-DXA parameter. **(A)** Horizon Wi, 251 HAF Hospital relative to Hospital del Mar. **(B)** Horizon A, Washington University relative to Hôpital Lapeyronie. For visualization, regression coefficients and 95% confidence intervals were standardized by the standard deviation of absolute scan-rescan error within each parameter and scanner-model subset. Points indicate standardized adjusted center differences, and horizontal lines indicate standardized 95% confidence intervals. Raw ($p$) and Benjamini-Hochberg-adjusted ($p_{adj}$) $p$-values are shown for each parameter, corrected within each panel's family of five tests.